\documentclass[supp]{new_tlp} %% added supp option
\usepackage{supp} %% added
\usepackage{mathptmx}
\usepackage{verbatim}
\usepackage{amssymb}  %% \mathbb
\usepackage{epsfig}
\usepackage{subfigure}
\usepackage[justification=centering]{caption}
\usepackage[noend]{algorithmic}
\usepackage{algorithm2e}
\usepackage{textcomp}
\let\proof\relax

\usepackage{amsthm} % definition of example and definition
\theoremstyle{definition}
\newtheorem{definition}{Definition}[section]
\let\olddefinition\definition
\renewcommand{\definition}{\olddefinition\itshape}
\newtheorem{example}{Example}[section]
\newcommand\bcmdtab{\noindent\bgroup\tabcolsep=0pt%
  \begin{tabular}{@{}p{10pc}@{}p{20pc}@{}}}
\newcommand\ecmdtab{\end{tabular}\egroup}

  \title[CDF-Intervals]
        {CDF-Intervals: A Reliable Framework to Reason about Data with Uncertainty}
  \author[A. SAAD]
         {AYA SAAD\\
         Universit\"{a}t Ulm, Germany\\
         \email{{ayas}@aucegypt.edu}
         }
\jdate{March 2014}
\pubyear{2014}
\pubauthor{Saad} %% added, please replace "YourLastName" appropriately
\pagerange{\pageref{firstpage}--\pageref{lastpage}}
\begin{document}
\label{firstpage}
\maketitle
  \begin{abstract}
This research introduces a new constraint domain for reasoning about data with uncertainty. It extends convex modeling with the notion of p-box to gain additional quantifiable information on the data whereabouts. Unlike existing approaches, the p-box envelops an unknown probability instead of approximating its representation. The p-box bounds are uniform cumulative distribution functions ({\em cdf}) in order to employ linear computations in the probabilistic domain. The reasoning by means of p-box {\em cdf}-intervals is an interval computation which is exerted on the real domain then it is projected onto the {\em cdf} domain. This operation conveys additional knowledge represented by the obtained probabilistic bounds. Empirical evaluation shows that, with minimal overhead, the output solution set realizes a full enclosure of the data along with tighter bounds on its probabilistic distributions.
  \end{abstract}
\section{Introduction}
This research proposes a novel constraint domain for reasoning about data with uncertainty. The work was driven by the practical usage of reliable approaches in Constraint Programming (CP). These approaches tackle large scale constraint optimization (LSCO) problems associated with data uncertainty in an intuitive and tractable manner. Yet they have a lack of knowledge when the data whereabouts are to be considered. These whereabouts often indicate the data likelihood or chance of occurence, which in turn, can be ill-defined or have a fluctuating nature. It is important to know the source and type of the data whereabouts in order to reason about them. The purpose of this novel framework is to intuitively describe data coupled with uncertainty or following unknown distributions without losing any knowledge given in the problem definition. The p-box {\em cdf}-intervals extend the {\em cdf}-intervals approach, \cite{saad2010constraint}, with a p-box structure to obtain a safe enclosure. This enclosure envelops the data along with its whereabouts with two distinct quantile values, each is issuing a {\em cdf}-uniform distribution, \cite{saadcdf}. 
This research is concerned with the following contributions: (1) a new uncertain data representation specified by p-box {\em cdf}-intervals, (2) a constraint reasoning framework that is used to prune variable domains in a p-box {\em cdf}-interval constraint relation to ensure their local consistency, (3) an experimental evaluation, using the inventory management problem, to compare the novel framework with existing approaches in terms of expressiveness and tractability. The expressiveness, in this comparison, measures the ability to model the uncertainty provided in the original problem, and the impact of this representation on the solution set realized. On the other hand, the tractability measures the system time performance and scalability. The experimental work shows how this novel domain representation yields more informed results, while remaining computationally effective and competitive with previous work. 
\section{Preliminaries}
Models tackling uncertainty are classified under the set of plausibility measures \cite{halpern2003reasoning}. They are categorized as: possibilistic and probabilistic. Convex models, found in the world of {\em fuzzy} and interval/robust programming, are favored when ignorance takes place. They are adopted in the CP paradigm in {\em fuzzy} Constraint Satisfaction Problems (CSPs) \cite{dubois1996possibility} and numerical CSPs \cite{benhamou1995Interval}. Probabilistic models are best adopted when the data has a fluctuating nature. They are the heart of the probabilistic CP modeling found in valued CSP \cite{schiex1995}, semirings \cite{bistarelli1999}, stochastic CSPs \cite{walsh2008}, scenario-based CSPs \cite{tarim2006} and mixed CSPs \cite{fargier1996}.
Techniques adopting convex modeling are characterized to be more conservative. They can often consider many unnecessary outcomes along with important ones. This conservative property supplements convex modeling with a high tractible and scalable behavior since operations, in these models, are exerted on the convex bounds only. On the other hand, probabilistic approaches add a quantitative information that expresses the likelihood, yet these approaches impose assumptions on the distribution shape in order to conceptually deal with it in a mathematical manner. Moreover, probabilistic mathematical computations are very expensive because they often depend on the non-linear probability shape. The research objective is to introduce a novel framework: the p-box {\em cdf}-intervals. It is based on a probability box (p-box) structure \cite{ferson2003constructing} that envelops a set of cumulative distribution functions ({\em cdf}). The p-box concept is adopted in the literature, specifically when the environment is uncertain, to represent an unknown distribution with a safe enclosure rather than depending on statistical approximation. A {\em cdf} is a monotone (non-decreasing) function that  indicates for a given uncertain value the probability that the
actual data lies before. It defines the aggregated probability of a value to occur. The p-box bounding {\em cdf} distributions in the proposed framework are uniform, each is represented by a line equation in order to maintain an inexpensive computational complexity. The key idea behind the construction of the p-box {\em cdf}-intervals is to combine techniques from the convex models, to take advantage of their tractability, with approaches revealing quantifiable information from the probabilistic and stochastic world, to take advantage of their expressiveness. 

The framework is based on CP concepts because they proved to have a considerable flexibility in formulating real-world combinatorial problems. The CP paradigm aims at building descriptive algebraic structures which are easily embedded into declarative programming languages. These structures are heavily used in problem solving environment by  specifying conditions that need to be satisfied and allow the solver to search for feasible solutions. The following section demonstrates how to intuively represent the uncertainty, already given in the problem definition, in order to reason about it by means of the p-box {\em cdf}-intervals. A comparison between the novel representation of the data uncertainty with existing possibilistic and probabilistic approaches is also taking place in order to demonstrate the model expressiveness. This representation is input to the solver with a new domain specification. Consequently the reasoning about this new specification is defined. It proves how the reasoning by means of p-box {\em cdf}-intervals is tractable. Accordingly, combining reasoning techniques from convex models with quantifiable information from probabilistic models yields a novel model that is together tractable and expressive. 
\section{Input Data Representation}\label{sec:datarepresentation}
\begin{figure}[h]
\centering
\subfigure[]{\includegraphics[scale=0.16]{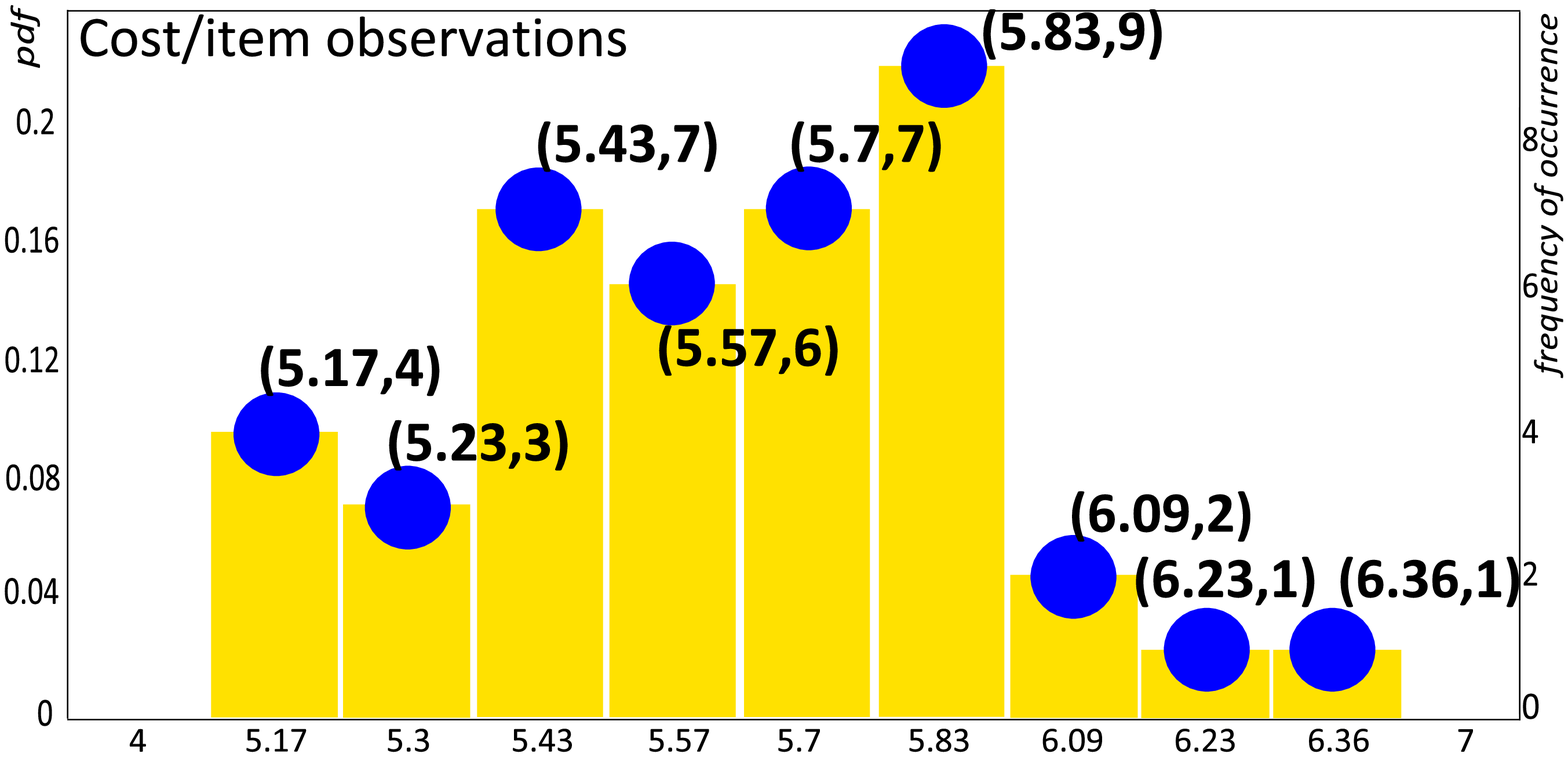}}
\subfigure[]{\includegraphics[scale=0.16]{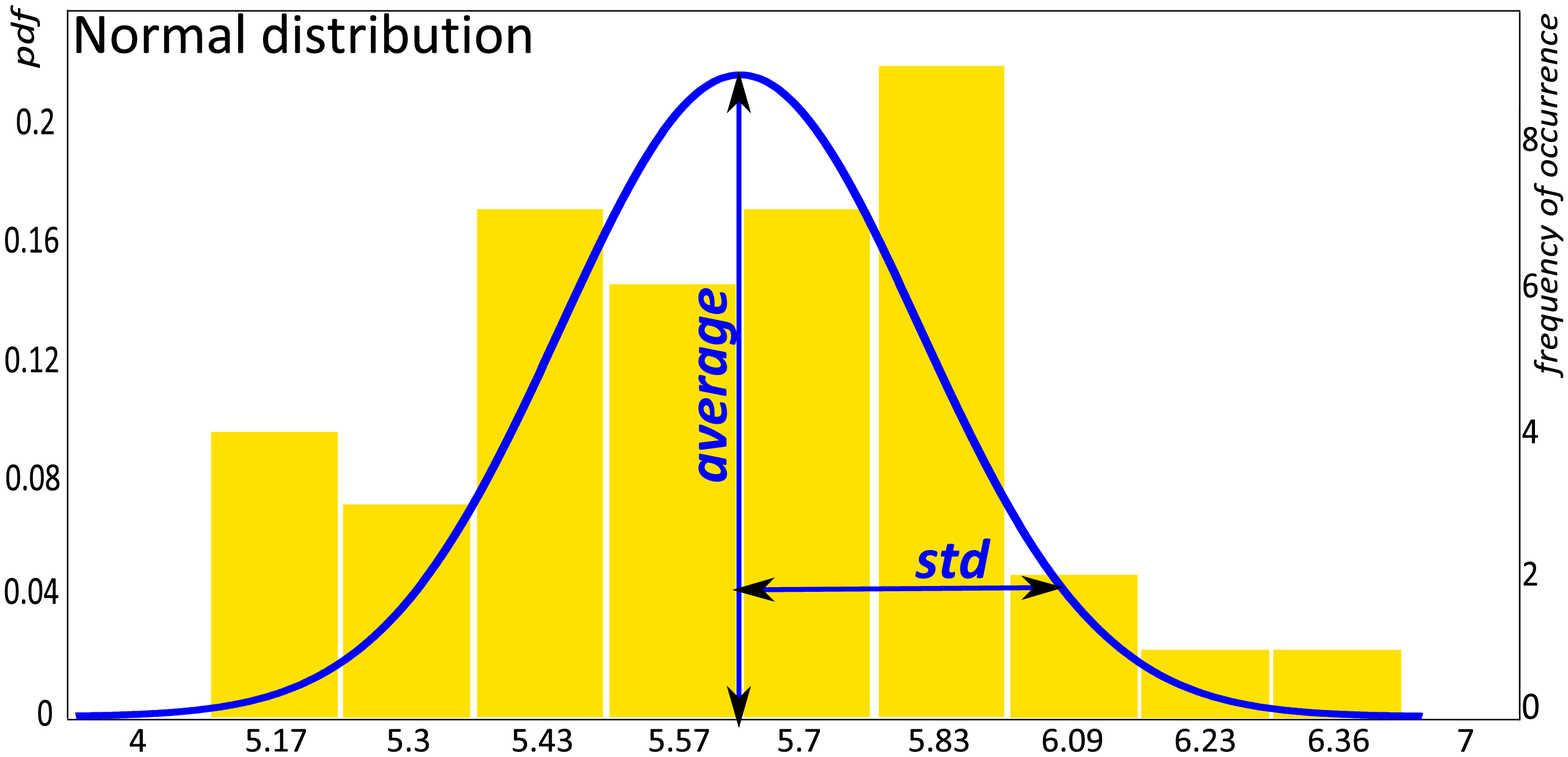}}
\subfigure[]{\includegraphics[scale=0.16]{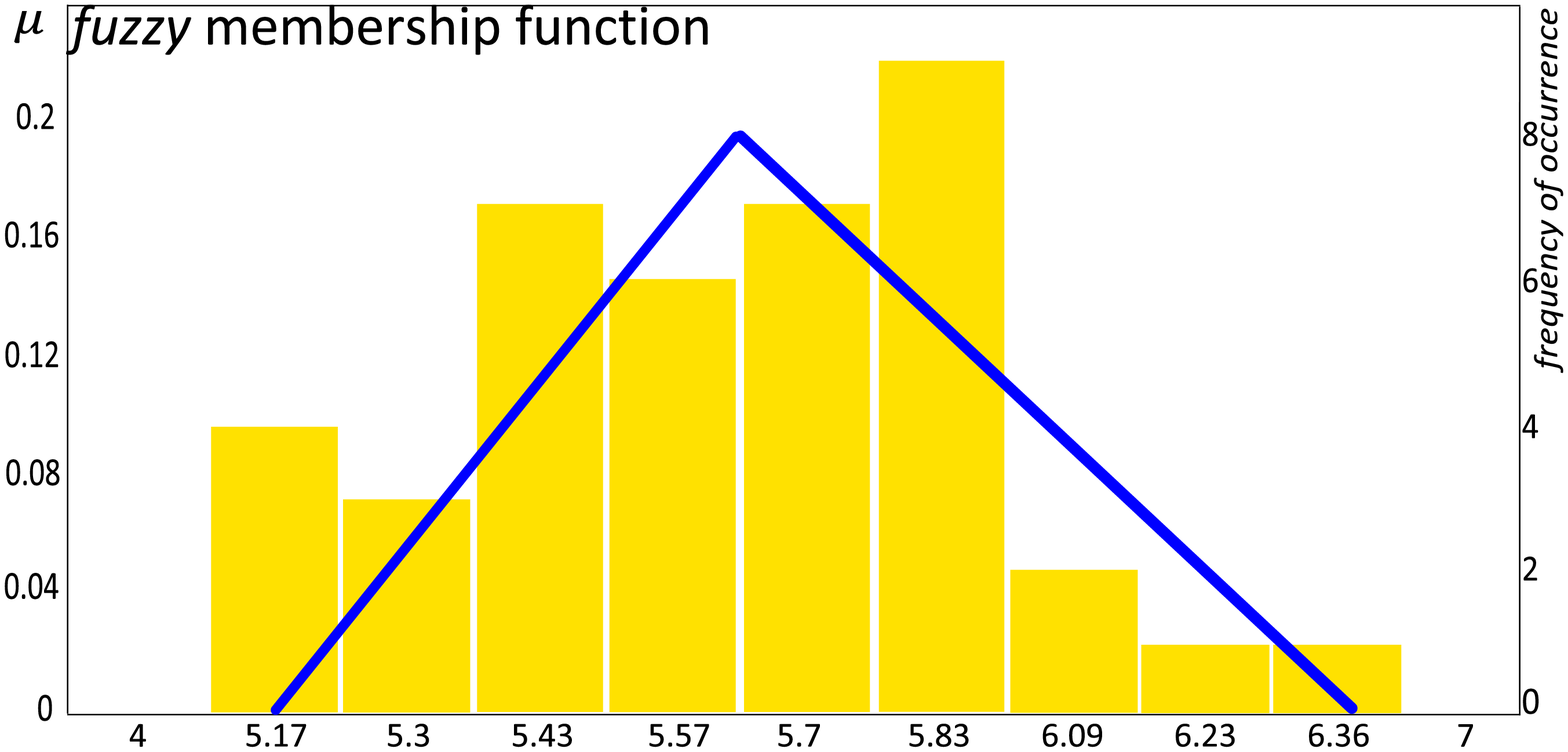}}
\vspace{-1\baselineskip}
\caption{Varying cost of the steel stud item and its probability histogram: (a) genuine observations (b) Normal distribution (c) {\em fuzzy} distribution} \label{fig:costperitemobservationshistogram}
\end{figure}

\begin{figure}[h]
\centering
\subfigure[]{\includegraphics[scale=0.16]{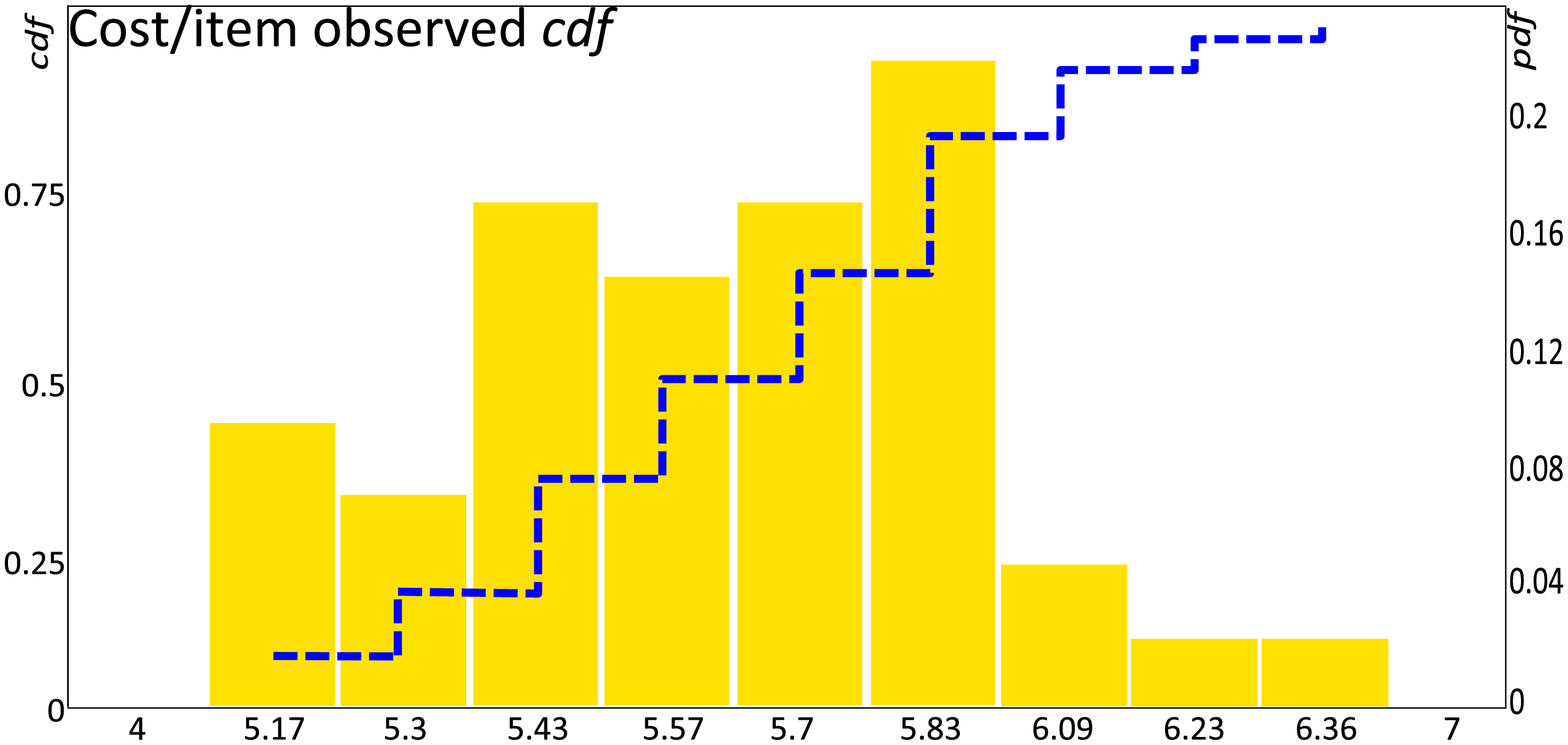}}
\subfigure[]{\includegraphics[scale=0.16]{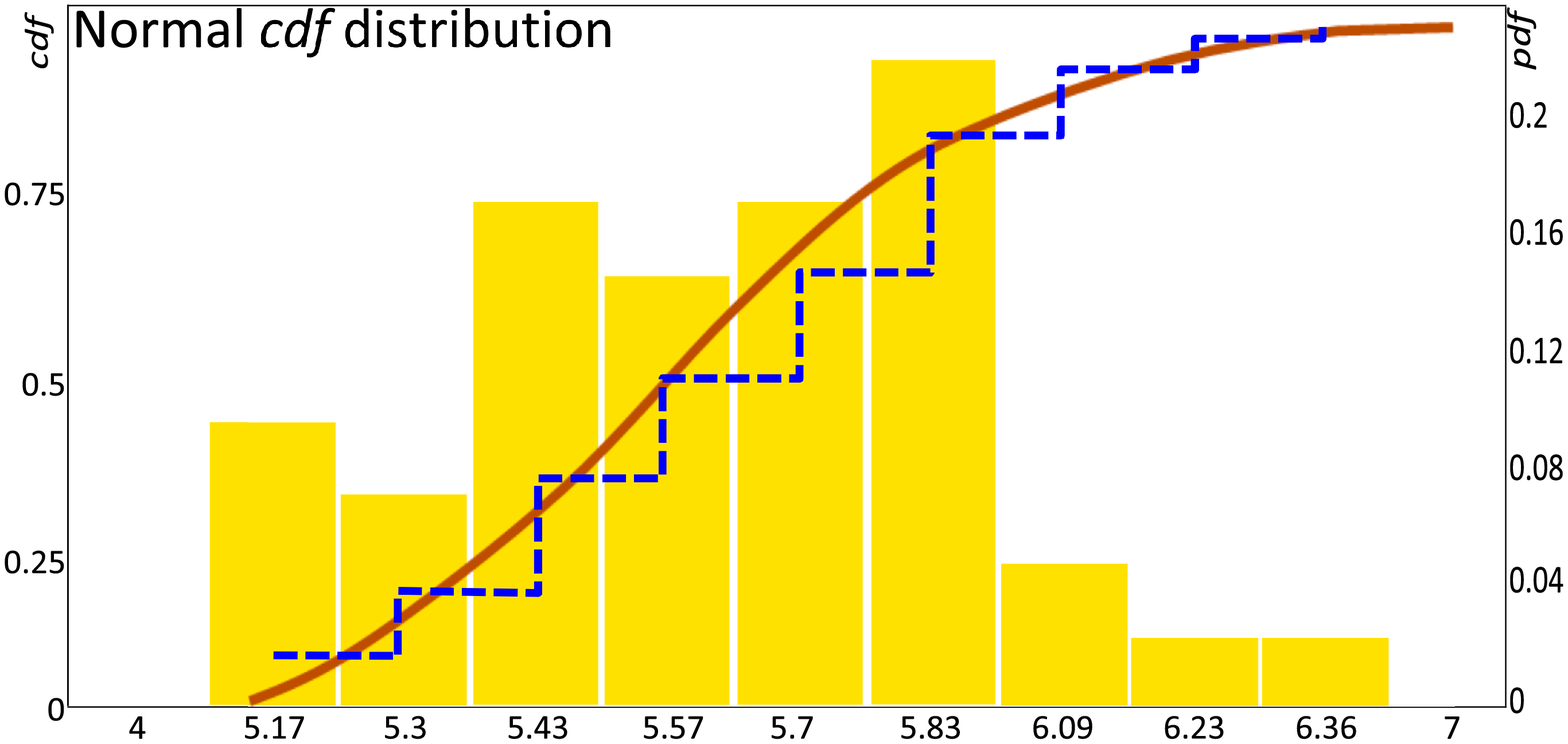}}
\subfigure[]{\includegraphics[scale=0.16]{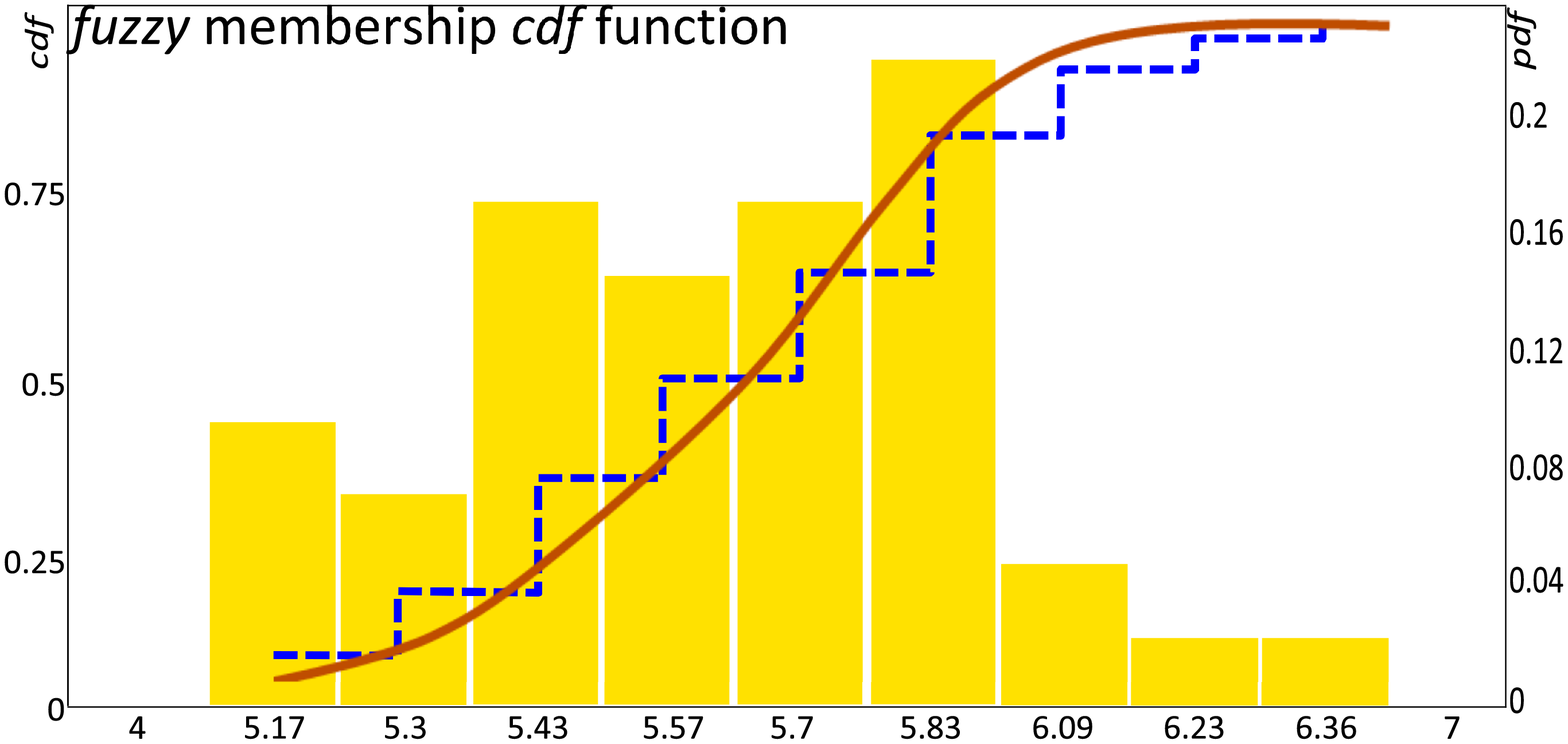}}\\
\subfigure[]{\includegraphics[scale=0.16]{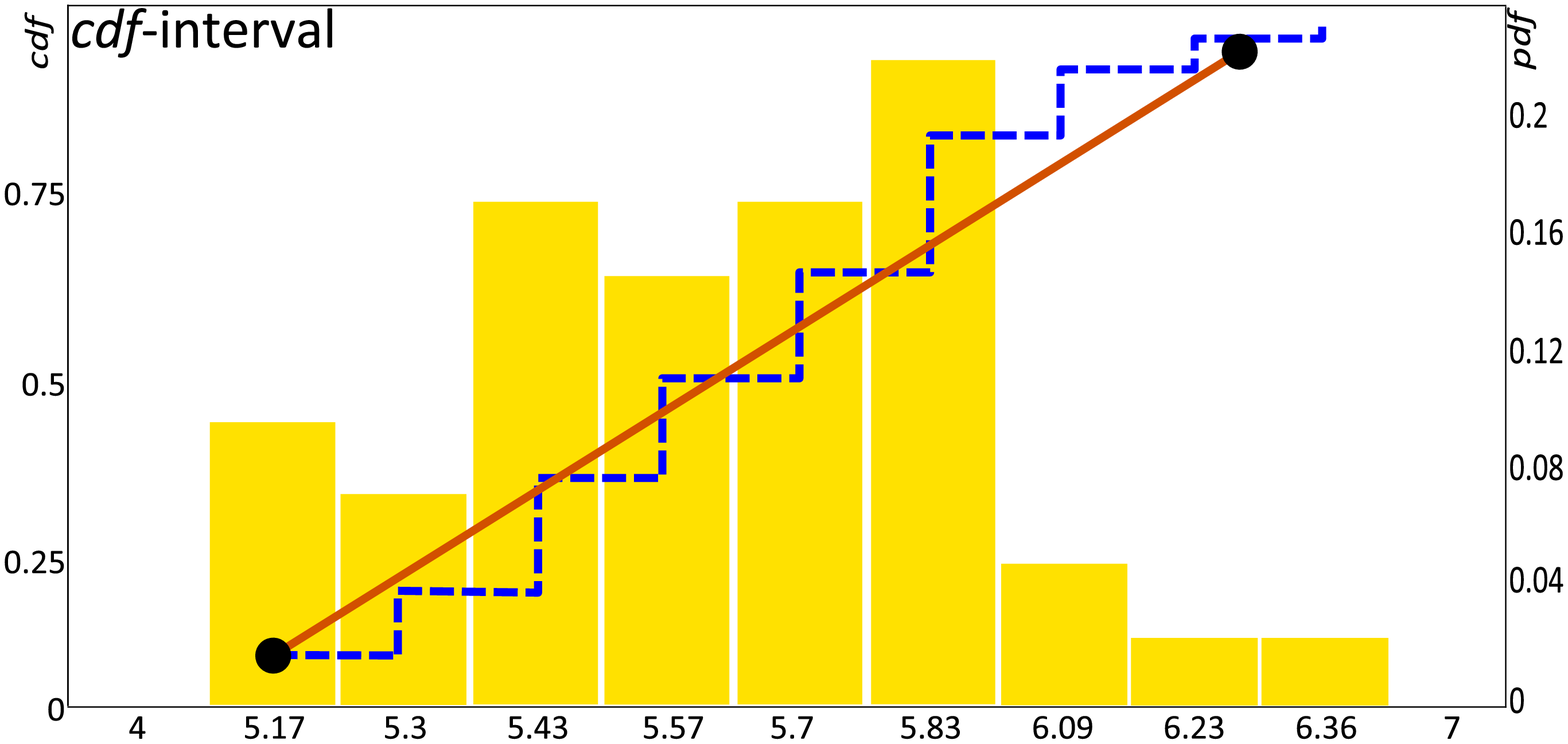}}
\subfigure[]{\includegraphics[scale=0.16]{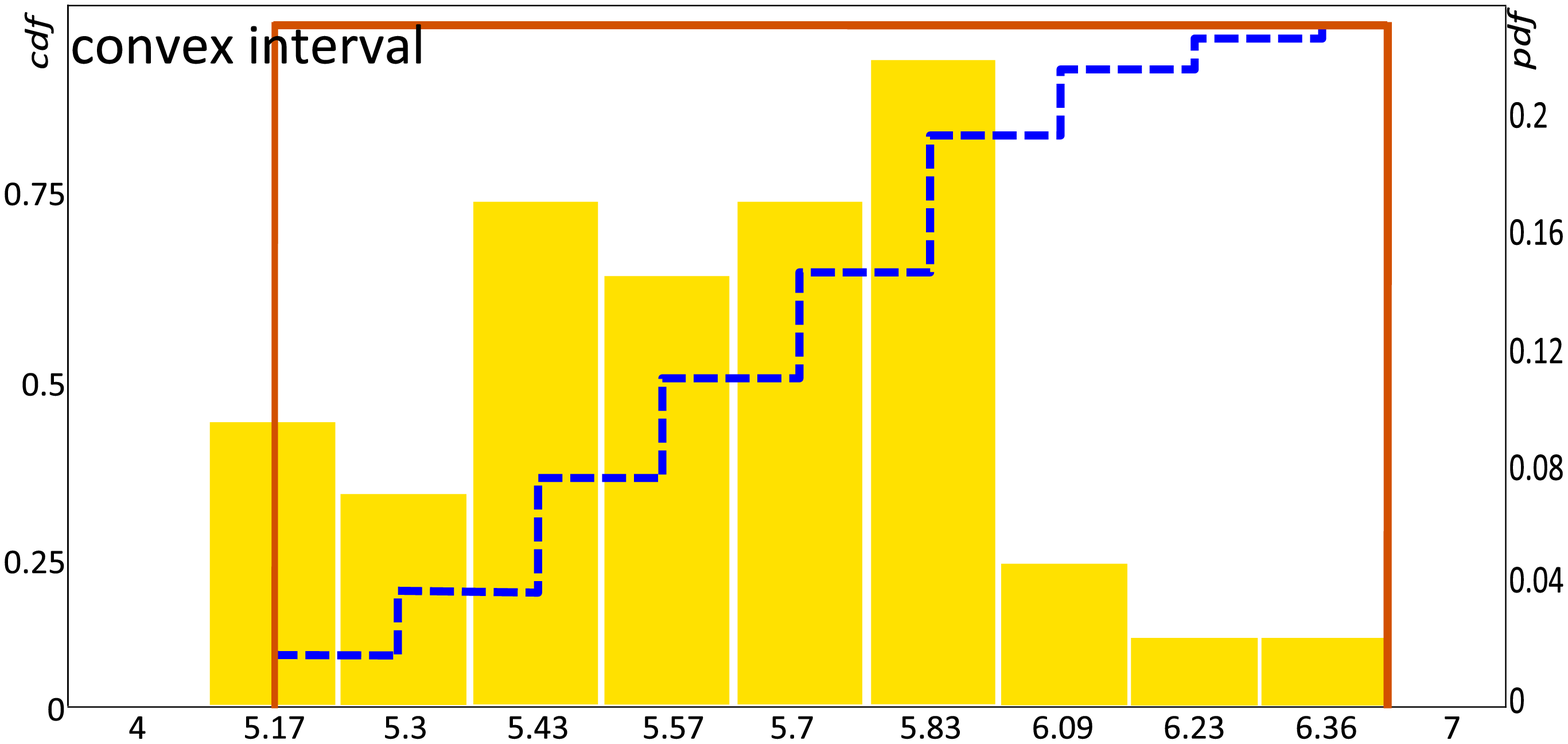}}
\subfigure[]{\includegraphics[scale=0.16]{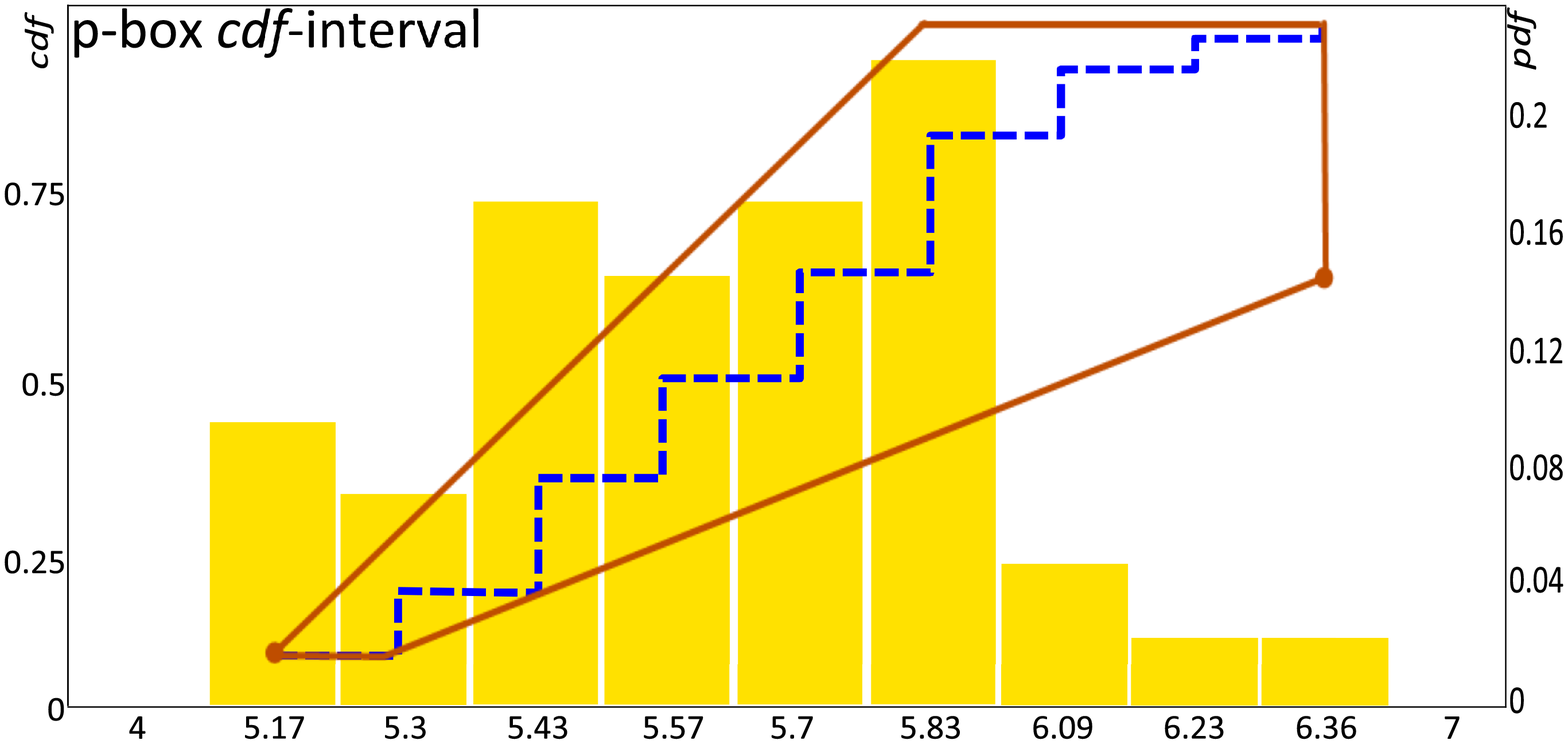}}
\vspace{-1\baselineskip}
\caption{Varying cost of the steel stud item, its derived probabilities, and its {\em cdf} distributions} \label{fig:cdfprojection}
\end{figure}

Quantifiable information is often available during the data collection process, but lost during the reasoning because it is not accounted for in the representation of the uncertain data. This information however is crucial to the reasoning process, and the lack of its interpretation yields erroneous reasoning because of its absence in the produced solution set. It is always necessary to quantify uncertainty that is naturally given in the problem definition in order to obtain robust and reliable solutions.
\begin{example}\label{ex:costperitemdataobservation}
Consider, as a running example, the varying cost observations of a steel stud manufacturing item. Fig. \ref{fig:costperitemobservationshistogram}(a) illustrates the cost variations along with their corresponding frequencies of occurrence. For instance, the point $(5.17,4)$ is the amount of the cost/item, equal to $5.17$, and observed $4$ times during the past $2$ years (corresponding to a population $m = 40$). $9$ is the number of distinct measured quantiles. The minimum and the maximum observed values, in this example, are $5.17$ and $6.36$ respectively. 
\end{example}
To compute the probabilistic/ possibilistic representations, the average and standard deviation of the observed population are derived. In this example, they are equal to $5.6$ and $0.28$ respectively. The nearest Normal probability distribution and the {\em fuzzy} membership function are illustrated in Fig. \ref{fig:costperitemobservationshistogram} (b) and (c). 
To compare the data representations adopted in various approaches, the observed data is projected onto the {\em cdf}-domain. By definition, the {\em cdf} is a monotonic distribution that keeps the probabilistic information in an aggregated manner. Information obtained from the measurement process is often discrete and incomplete, hence, its {\em cdf}-domain projection forms a staircase shape \cite{smith1992approximation}. The {\em cdf} distribution of the genuine observed data whereabouts is depicted in the running example by the dotted staircase shape in Figure \ref{fig:cdfprojection}. Normal and {\em fuzzy} {\em cdf} distributions are shown by the continuous red curves in Fig. \ref{fig:cdfprojection} (b) and (c). Each is based on an approximation that lacks precise point fitting of the original data whereabouts. Similarly, the {\em cdf}-interval, in Fig. \ref{fig:cdfprojection} (d), approximates the data whereabouts by means of a line connecting the two bounding data values. The convex model representation however shapes a rectangle, illustrated in Fig. \ref{fig:cdfprojection} (e). This rectangle includes all values in the {\em cdf} range $[0,1]$. The convex representation treats data values lying within the interval bounds equally, i.e. it lacks the probabilistic information. The p-box {\em cdf}-interval, depicted in Fig. \ref{fig:cdfprojection} (f) enforces tighter bounds on the probabilities when compared to convex models illustrated in Fig. \ref{fig:cdfprojection} (e). This envelopment guarantees a safe enclosure on the unknown distribution while preserving tractability due to the fact that its bounds are represented each by a line equation.
\paragraph{Interpretation of the p-box cdf confidence interval ${\bf I}$.}
For a given interval of points specified by ${\bf I}=[p_a,p_b]$, $p_a$ and $p_b$ are the extreme points which bound the p-box {\em cdf}-interval.  
One can see that this interval approach does not aim at approximating
the curve but rather enclosing it in a reliable manner. The complete envelopment is exerted by means of the uniform {\em cdf}-bounds, which are depicted by the red curves in Fig. \ref{fig:cdfprojection} (f). It is impossible to find a point that exists outside the formed interval bounds. The {\em cdf} bounds are chosen to have a uniform distribution. Each is represented by a line with a slope issued from one of the extreme quantiles. Storing the full information of each bound is sufficient to restore the designated interval assignment. Bounds are denoted by triplet points, in the $2$D space, to guarantee the full information on: the extreme quantile value observed; the {\em cdf}-line issued from this observed value; and the degree of steepness formed by this line. The slope of the uniform {\em cdf}-distribution indicates how the probabilistic values accumulate for successive quantiles on the line. Accordingly, the p-box {\em cdf}-interval point representation: $p_a = (a,F^p_a,S^p_a)$ and $q_b = (b,F^q_b,S^q_b)$. 
\begin{definition}
$S^p_x$ is the slope of a given cdf-distribution; it signifies the average step probabilistic value. For a given uniform cdf-distribution 
\begin{equation}
S^p_x = \frac{{F_b - F_a}}{b - a}, \forall a \leq x \leq b
\end{equation} 
\end{definition}
Plotting a point $p_x$ within the p-box {\em cdf}-interval deduces bounds on its possible chances of occurrence.
\begin{definition}\label{def:FPBOXpx} $F_x^I$ is the interval of values obtained when $p_x$ is projected onto the {\em p-box} cdf bounds. For a point $p_x \in {\bf I}$ denoted as $p_x=(x, F_x^p, S_x^p)$ 
\begin{equation}
a<x<b,~\mbox{and}~ F_b^{q`} \geq F_x^I \geq F_a^{p`} ~\mbox{and}~ S^p_a \geq S^p_x \geq S^q_b
\end{equation} 
\end{definition}
$F^{p`}_a$ and $F^{q`}_b$ are the possible maximum and minimum {\em cdf} values $p_x$ can take; both are computed by projecting the point $p_x$ onto the {\em cdf} distributions passing through real points $a$ and $b$ respectively. They are derived using the following linear projections, computed in $O(1)$ complexity:
\begin{equation}
F_a^{p`} = min(S_a^p(x-a) + F_a^p,1) ~~~~ \mbox{and} ~~~~ F_b^{p`} = max(F_b^p - S_b^p(b-x),0) \nonumber
\end{equation}
The equation above guarantees the probabilistic feature of the {\em cdf}-function by restricting its aggregated value from exceeding the value $1$ and having negative values below $0$.
\begin{example}
${\bf I} = [(5.17, 0.1, 1.2), (6.36, 0.7, 0.57)]$ is the p-box {\em cdf}-interval of the cost/item in Example \ref{ex:costperitemdataobservation}. Suppose that $x_i=5.5$, its {\em cdf}-bound values $F_x^I = [0.2,0.5]$. This means that the possible chance of the value to be at most $5.5$ is between $20\%$ and $50\%$, with an average step probabilistic value between $0.57$ and $1.2$. Note that this interval is opposed to only one approximated value $F_x = 0.37$ in the {\em cdf}-intervals representation proposed in \cite{saad2010constraint}, the {\em fuzzy} {\em cdf} value $F_x = 0.31$ and its Normal {\em cdf} value is $F_x = 0.42$. Note that convex models do not enforce any probabilistic bounds, accordingly, $x_i=5.5$ has a {\em cdf} $F^I_x \in [0,1]$.  
\end{example}
\section{Constraint reasoning}\label{sec:constraintreasoning}
In the CP paradigm, relations between variables are specified as constraints. A set of rules and algebraic semantics, defined over the list of constraints, formalize the reasoning about the problem. As a fundamental language component in the Constraint Logic Programming (CLP), these set of rules, with a syntax of definite clauses, form the language scheme \cite{jaffar1987}. The constraint solving scheme is intuitively and efficiently utilized in the reasoning over the computation domain. The scheme formally attempts at assigning to variables a suitable domain of discourse equipped with an equality theory together with a least and a greatest model of fix-point semantics. Starting from an initial state the reasoning scheme follows a local consistency technique which attempts at constraining each variable over the p-box {\em cdf}-interval domain while excluding values which do not belong to the feasible solution. An implementation of the constraint system was established as a separate module in the ECL{\em $^i$}PS{\em $^e$} constraint programming environment \cite{eclipse}. ECL{\em $^i$}PS{\em $^e$}  provides two major components to build the solver: an attributed variable data structure and a suspension handling mechanism. Fundamentally, attributed variables are specific data structures which attach more than one data type. Together they permit for a new definition of unification which extends the well-known Prolog unification \cite{lehuitouze1990,holzbaur92}. A p-box {\em cdf}-interval point is implemented in an attributed variable data structure with three main components: quantile, {\em cdf} value and slope. Whilst constraints suspension handling is a highly flexible mechanism that aims at controlling user defined atomic goals. This is achieved by waiting for user-defined conditions to trigger specific goals.
 
\begin{figure}[hc]
\centering
\subfigure[]{\includegraphics[scale=0.2]{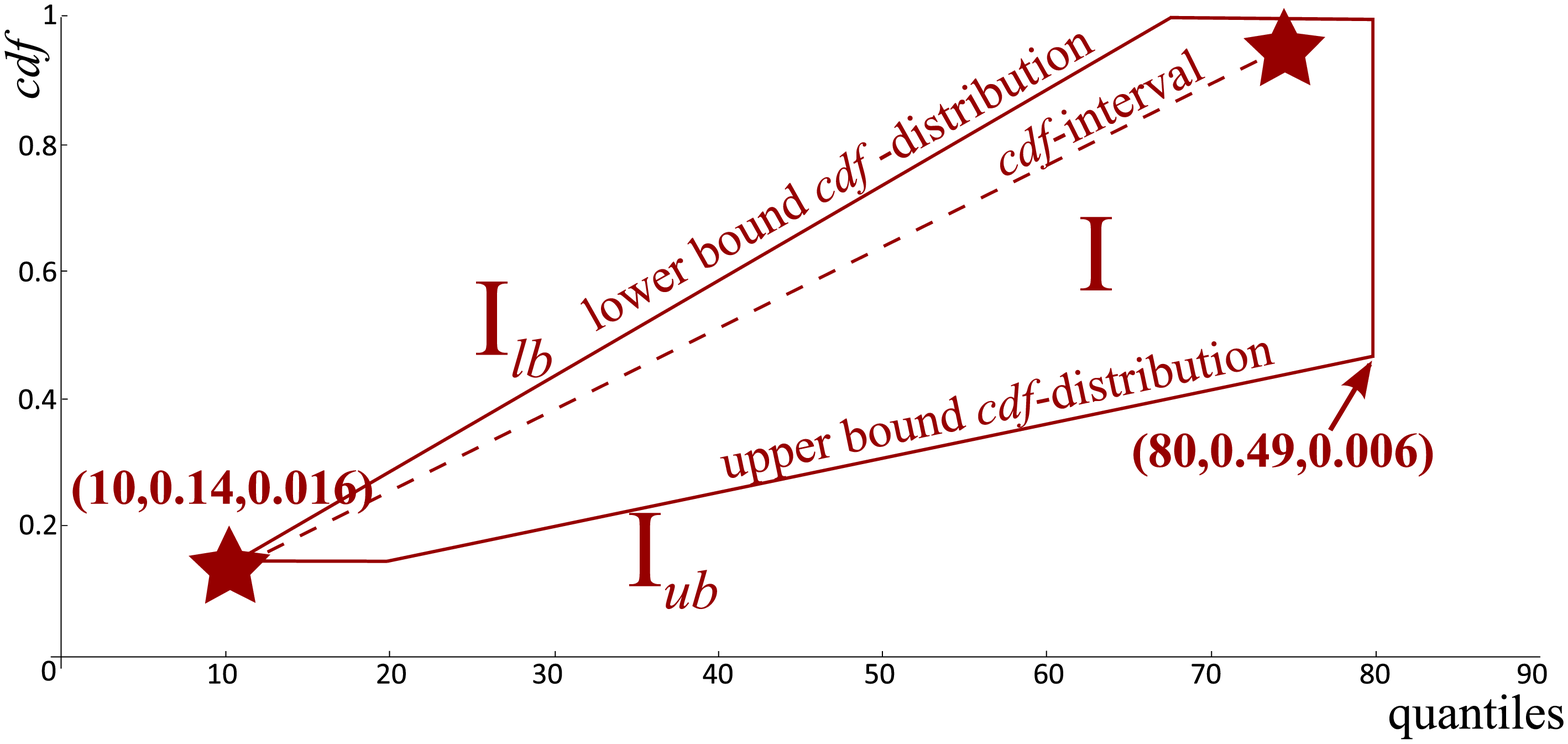}}
\subfigure[]{\includegraphics[scale=0.2]{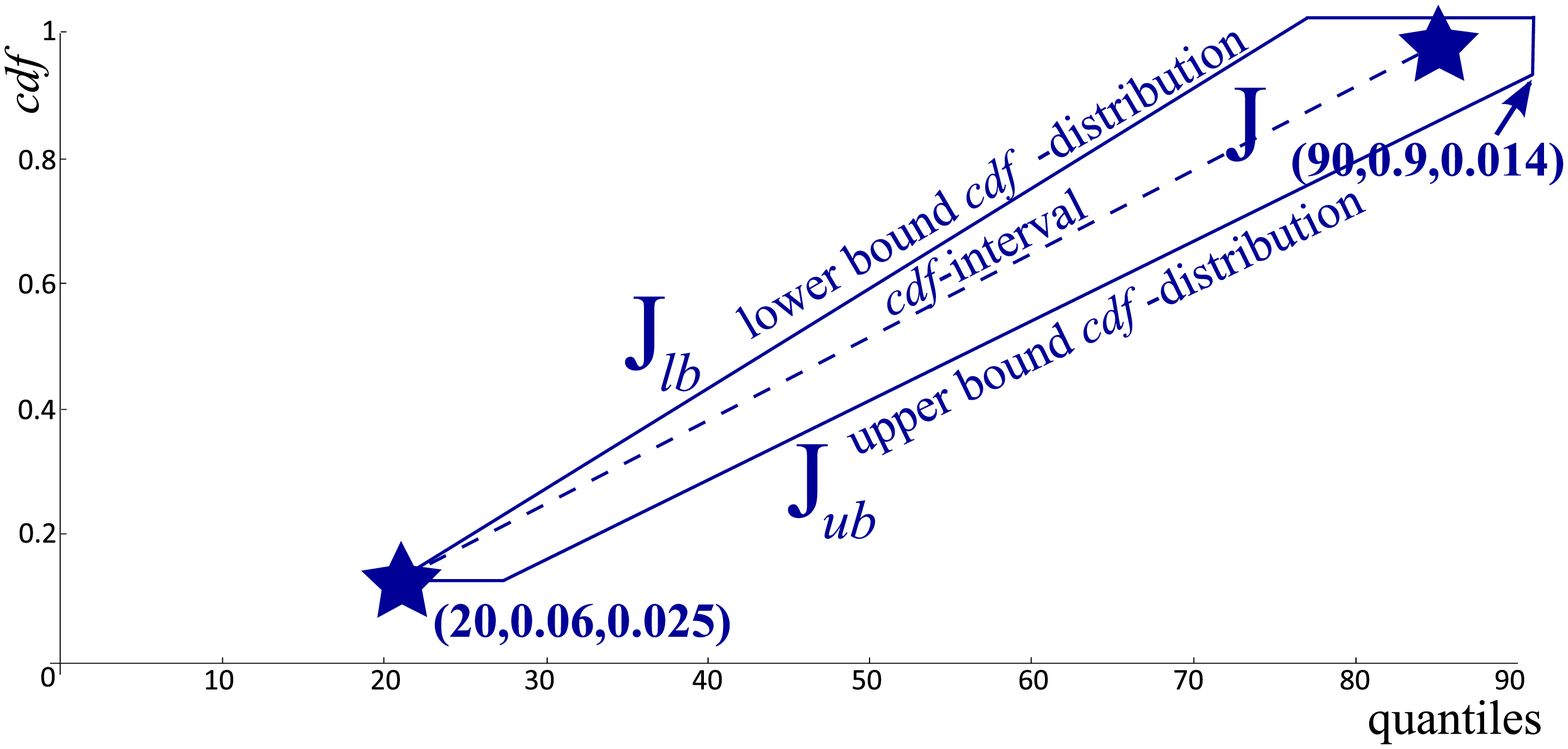}}\\
\subfigure[]{\includegraphics[scale=0.2]{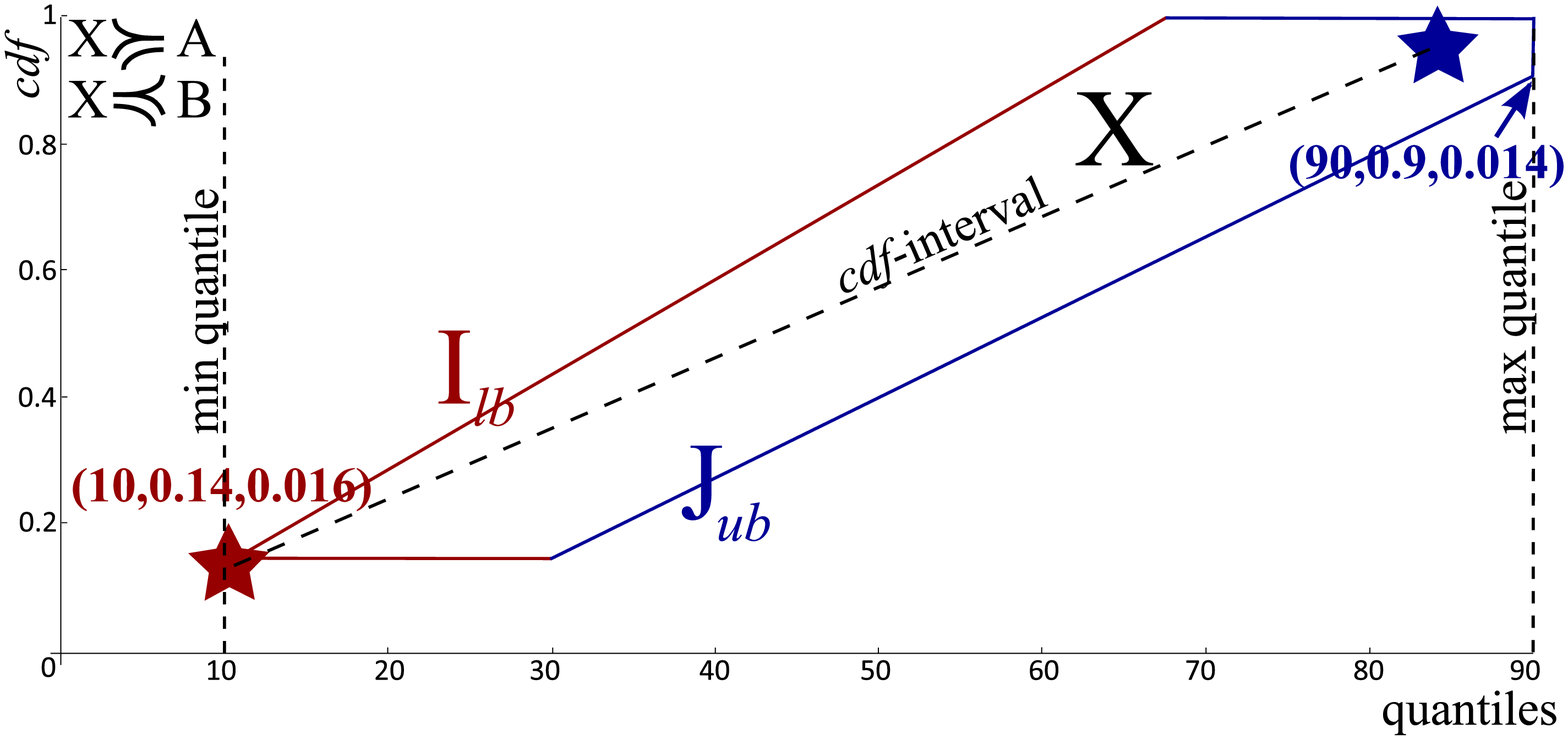}}
\subfigure[]{\includegraphics[scale=0.2]{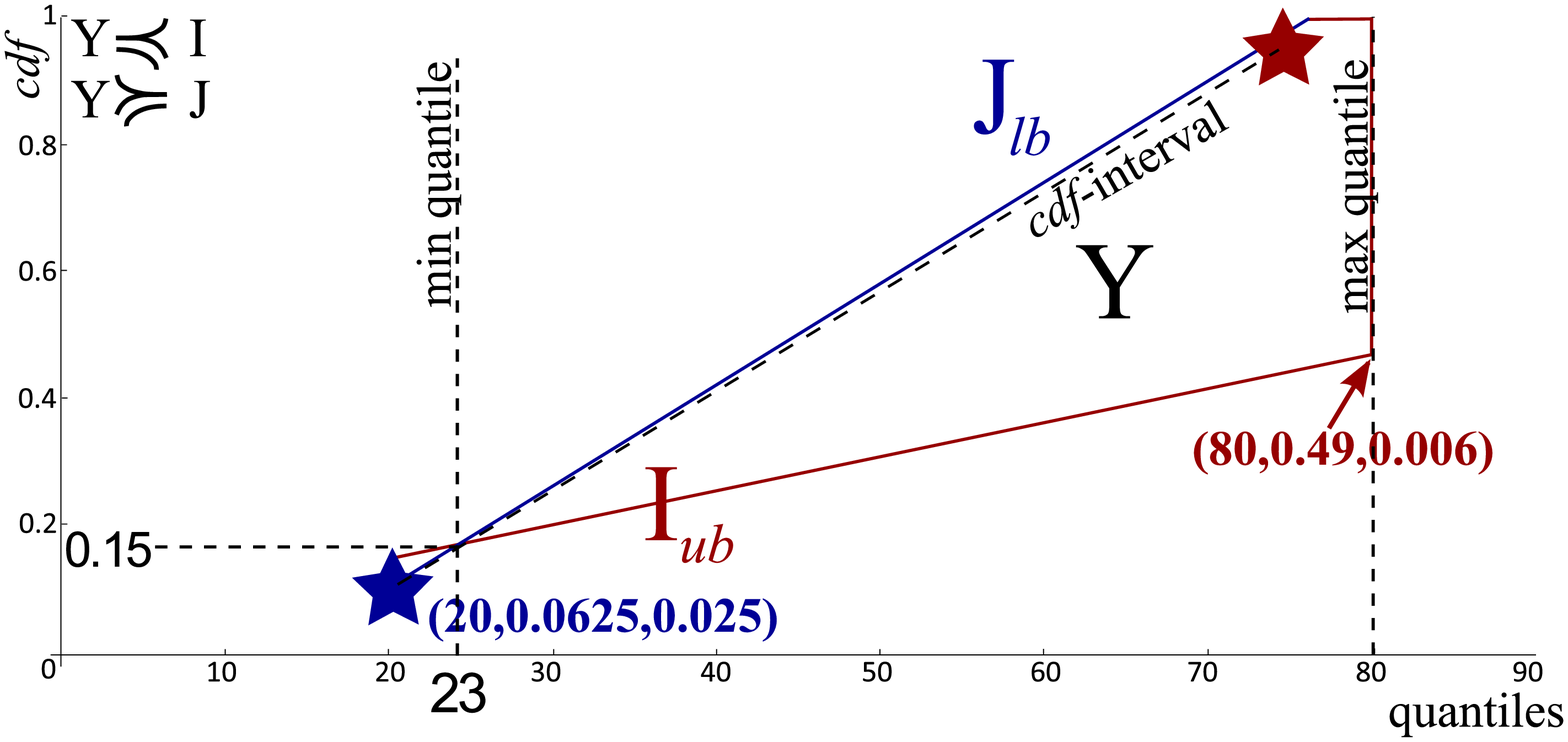}}
\vspace{-1\baselineskip}
\caption{Ordering constraint execution} \label{fig:orderingconstraint}
\end{figure}

Implemented rules in our solver infer the local consistency in the p-box {\em cdf}-interval domains of the binary equality and ordering constraints $\{ =, \preccurlyeq_\mathcal{U} \}$, and that of the ternary arithmetic constraints $\{ +_\mathcal{U}, -_\mathcal{U}, \times_\mathcal{U}, \div_\mathcal{U} \}$. Operations, in the solver, are exerted first as real interval computations, and then they are projected onto the {\em cdf} domain using a linear computation, as shown in Definition \ref{def:FPBOXpx}. This section demonstrates how the ordering and the ternary addition constraints infer the local consistency over the variable domains of $X$, $Y$, and $Z$ assuming that their initial bindings are $I = [p_a,p_b]$, $J = [q_c,q_d]$ and $K = [r_e,r_f]$ respectively. The ternary multiplication, subtraction and division constraints are implemented in the same way.
\paragraph{Ordering constraint $X \preccurlyeq_{\mathcal{U}} Y$}. To infer the local consistency of the binary ordering constraint, the lower {\em cdf}-bound of $X$ is extended and the upper {\em cdf}-bound of $Y$ is contracted. 
\begin{example}
Let ${\bf I}$ and ${\bf J}$ be two p-box {\em cdf}-interval domains. ${\bf I} = [(10,0.14,0.016),(80,0.49,0.06)]$ and ${\bf J} = [(20,0.06,0.025),(90,0.9,0.014)]$. The effect of applying the set of constraints $X \succcurlyeq_{\mathcal{U}} {\bf I}$ and $X \preccurlyeq_{\mathcal{U}} {\bf J}$, prunes the domain of $X$. As a result, the variable $X$ is bounded by the lower bound of ${\bf I}$ and by the upper bound of ${\bf J}$: $X \in [(10,0.14,0.016),(90,0.9,0.014)]$ as shown in Fig. \ref{fig:orderingconstraint} (a). Clearly the obtained domain of $X$, in this example, preserves the convex property of the p-box {\em cdf}-intervals. 
Let $Y$ be subject to the domain pruning using the set of constraints: $Y \preccurlyeq_{\mathcal{U}} {\bf I}$ and $Y \succcurlyeq_{\mathcal{U}} {\bf J}$. As a result, $Y$ should be bounded by the lower bound of $\bf{J}$ and the upper bound of $\bf{I}$. However, in this case, at lower quantiles $\leq 23$, the upper bound distribution of $\bf{I}$ preceeds the lower bound of $\bf{J}$. The fact that conflicts the stochastic dominance property of a p-box {\em cdf}-interval domain. In order to resolve this conflict, the real bounds of $Y$ are further pruned to the point of the probability intersection $=23$. 
\end{example}
\paragraph{Ternary addition constraints $X +_\mathcal{U} Y = Z$}. The addition operation is implemented by summing up pair of points, defined in the $2$D space and located within the p-box {\em cdf}-interval bounds which enclose the domain ranges of $X$ and $Y$. This addition operation is linear. It is convex and can be computed from the end points of the domains involved in the addition. The p-box {\em cdf}-domain of $Z$ is updated to envelop all points defined in that range. 
\begin{figure}[hc]
\centering
\subfigure{\includegraphics[scale=0.45]{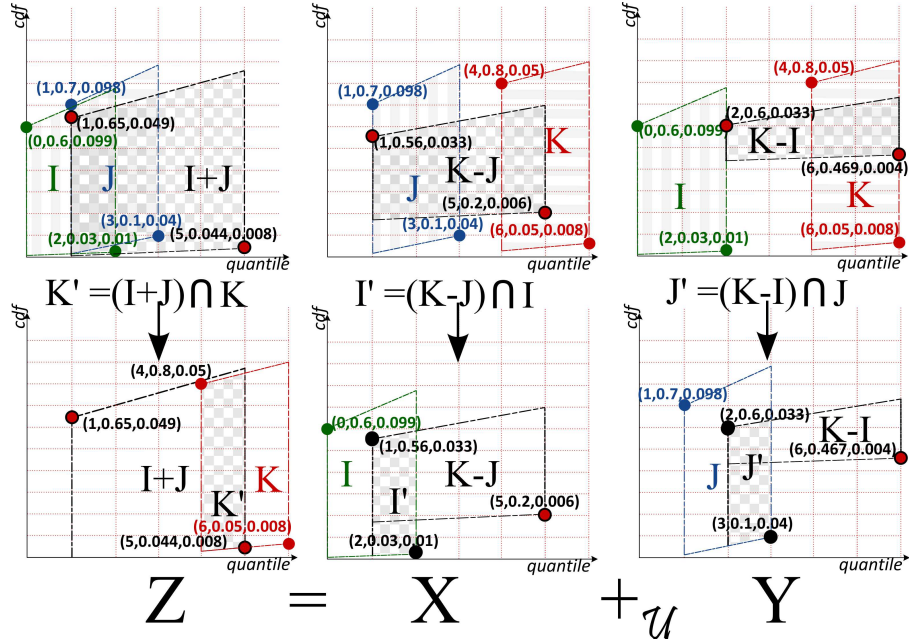}}
\vspace{-1\baselineskip}
\caption{Ternary addition inference rule execution: initial bindings are $X \in {\bf I}$,  $Y \in {\bf J}$ and  $Z \in {\bf K}$; final bindings are $X \in {\bf I}\textquotesingle$,  $Y \in {\bf J}\textquotesingle$ and  $Z \in {\bf K}\textquotesingle$.} \label{fig:ternaryaddition}
\end{figure}
\begin{example}
Fig. \ref{fig:ternaryaddition} depicts the execution steps of the p-box {\em cdf} ternary addition inference rule, exerted on the variable domains involved in the relation $Z = X +_\mathcal{U} Y$. Observe that domain pruning is performed in a $2$ dimensional manner: quantile and {\em cdf}. The addition of the two variables $X$ and $Y$ is performed on the bounds of their predefined domains then it is projected onto the initial bindings. The first row in Fig. \ref{fig:ternaryaddition} shows output domains from the addition $I+J$, $K - J$ and $K - I$. Domain operations are exerted on the extreme points. The second row illustrates the intersection of the output domains with the initial bindings, assigned to $Z$, $X$ and $Y$. Obtained domains from the ternary addition operation are ${\bf K}\textquotesingle $, ${\bf I}\textquotesingle $ and ${\bf J}\textquotesingle$. Clearly, in this example, pruning real quantile bounds is identical to that of real domains and since output domains preserve the stochastic dominance property no further pruning takes place. 
\end{example}
The ternary addition constraint exerted on p-box {\em cdf}-interval domains is a simple addition computation since it adopts the real-interval arithmetics which are then projected, linearly, onto the {\em cdf} domain. This operation is opposed to the {\em fuzzy} extended addition operation adopted in the constraint reasoning utilized in the possibilistic domain \cite{dutta2005single,petrovic1996fuzzy}, and to the Normal probabilistic addition which has a high computation complexity that is due to the Normal distribution shape \cite{glen2004computing}. 
\section{Empirical evaluation}\label{sec:empiricalevaluation}
The inventory management problem model proposed by \cite{tarim2004} is employed, as a case study, to evaluate the proposed framework. The key idea is to schedule ahead replenishement periods and find the optimal order sizes which achieve a minimum total manufacturing cost. A reorder point with order size $X_t$ should meet customer demands up to the next point of replenishment.
\begin{definition}\label{def:pboxcdfimodel}
An inventory management model defined over a time horizon of $N$ cycles is 
\begin{eqnarray}\label{eq:pboxcdfimodel}
\mbox{minimize   }~~~~TC = \Sigma^{N}_{t=1}(a\delta_t + hI_t + vX_t) \nonumber \\
\mbox{subject to    }~~~~\delta_t = \left\{ \begin{array}{cc}
1 & \mbox{if}~~X_t >0 \\
0 & \mbox{otherwise} \\
\end{array} \right\} \nonumber \\
I_t = I_0 + \Sigma^{t}_{i=1} ( X_i - d_i ) \nonumber \\
X_t, I_t \geqslant 0,~~ t = 0,1,...,N  
\end{eqnarray}
\end{definition}
The constituents of the total cost in the model are: the setup cost, holding cost and purchase cost. The setup cost is defined by the ordering cost multiplied by the number of times a replenishment takes place. The holding cost depends on the depreciation cost and the level of the inventory observed in a given cycle. The purchase cost is the reorder quantity multiplied by the varying cost/item. From this model, one can observe that all cost components are typically fluctuating and unpredictable especially in the real-life version of the problem. This is due to the unpredictability of customer demands and the variability of the cost/item. Accordingly, this model perfectly fits the purpose of the evaluation: comparing the behavior of the models when the environment is uncertain.    
\paragraph{Information realized in the solution set.}
The model is tested for a randomly distributed monthly demands over a time horizon $N = 10$ cycles. The p-box {\em cdf}-interval representation is constructed for each demand observation per cycle and for each observed varying cost component (ordering cost $a$, holding cost per item $h$ and varying cost per item $v$) to guarantee a safe enclosure on the data whereabouts. This is opposed to the {\em fuzzy} and probabilistic modeling which is based on the average demand values given in the set $d_t \in \{26,36,23,28,32,30,29,37,25,34\}$. The two later models set assumptions on the shape of the probability distribution adopted, as pointed out in Section \ref{sec:datarepresentation}. 
The solver executes the set of addition and equality constraints in the p-box {\em cdf}-interval domain. Constraints are triggered until stabilized and consistency is reached by means of the inference rules defined in Section \ref{sec:constraintreasoning}. The solver suggests $2$ to $5$ replenishment periods, with a total holding cost $[(8.5, 0.83, 4.4E-04),(137.98, 0.039, 7.5e-5)]$ and a total manufacturing cost $[(2739.6, 0.8, 3.3E-04), (6483.2, 0.03, 6.2e-5)]$. This output is opposed to $6$ replenishment periods realized by the {\em fuzzy} and the probabilistic models with a total holding cost $\$53.5$ and $\$52.05$ and a total manufacturing cost $\$3868.5$ and $\$3828.93$ respectively.
\begin{figure}[hc]
\centering
\subfigure{\includegraphics[scale=0.25]{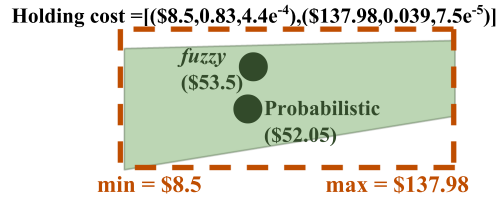}}
\vspace{-1\baselineskip}
\caption{Output solutions for holding cost} \label{fig:outputsolution}
\end{figure}
Fig. \ref{fig:outputsolution} illustrates a comparison between the output holding cost obtained from the models under consideration. The p-box {\em cdf}-interval graphical representation of the cost is depicted by the shaded region and their bounds in the convex models are illustrated by the dotted rectangles. Clearly, the solution set obtained from the p-box {\em cdf}-intervals model, when compared with the outcome of the convex model, realized an additional knowledge (i.e. tighter bounds in the {\em cdf} domain). This solution set is opposed to a one value proposed as $\$53.5$ by the {\em fuzzy} and as $\$52.05$ by the probabilistic models. Output solution point suggested by the latter models can, sometime, mislead or deviate the decision making. This is because their distributions are built, from the begining, on approximating the actual observed distribution.
\paragraph{Model tractability.}
We generate random distributions for monthly demands scaling up the problem time horizon for $\{7,10,24\}$ cycles. The first three rows in Table \ref{tab:memoryconsumption} show the real time taken by each model in seconds to generate the output solution of the total cost. Two other measurements, the shared heap used and the control stack used, are taken into consideration in order to study the memory consumption of each model. The shared heap used is the memory allocated to store compiled Prolog code and its related variables and necessary buffers. The control stack used is utilized to hold backtracking information. Table \ref{tab:memoryconsumption} demonstrates that stochastic model memory consumption grows exponentially when scaling-up the problem, it reaches $100\%$ of the memory usage for a time horizon $t=24$. The p-box {\em cdf}-intervals behavior is similar to convex models. Probabilistic and {\em fuzzy} models have the best shared heap utilization. Clearly the percentage of the control stack employed in the stochastic model is the highest. This is due to the behavior of the stochastic techniques which exhaustively build the solution scenarios in order to reach a solution. It is worth noting that convex models and p-box {\em cdf}-intervals do not need to build this tree since output solution set is provided within an interval range that is encapsulating all possible output scenarios. 
\begin{table}
\begin{minipage}{\textwidth}
\scriptsize 
\begin{tabular}{lrrrrrrrr}
\hline 
 & time horizon $t$ & stochastic & probabilistic & {\em fuzzy} & {\em cdf} & p-box & convex \\
\hline 
real time (sec)& $7$ & $0.43$ & $3.71$ & $3.06$ & $0.81$ & $0.78$ & $0.5$ \\
 & $10$ & $70.14$ & $6.08$ & $5.77$ & $3.28$ & $3.28$ & $3.06$ \\
 & $24$  & $1683.36$ & $175.22$ & $159.05$ & $55.41$ & $32.5$ & $31.2$ \\
\hline
shared heap used & $7$ & $0.21\%$ & $0.4\%$ & $0.29\%$ & $6.87\%$ & $6.86\%$ & $6.82\%$ \\
 & $10$ & $0.21\%$ & $0.5\%$ & $0.29\%$ & $9.68\%$ & $9.67\%$ & $9.62\%$ \\
 & $24$ & $100\%$ & $0.9\%$ & $0.7\%$ & $22.93\%$ & $23.04\%$ & $22.79\%$ \\
\hline 
control stack used & $7$ & $62.87\%$ & $46.71\%$ & $23.35\%$ & $0\%$ & $0\%$ & $0\%$ \\
 & $10$ & $89.82\%$ & $46.71\%$ & $23.35\%$ & $0\%$ & $0\%$ & $0\%$ \\
 & $24$ & $100\%$ & $46.71\%$ & $23.35\%$ & $0\%$ & $0\%$ & $0\%$ \\
\hline
\end{tabular}
\vspace{-1\baselineskip}
\caption{Real-time taken and measurement of memory consumption}
\label{tab:memoryconsumption}
\vspace{-2\baselineskip}
\end{minipage}
\end{table} 
Evidently, convex models outperfrom the rest of the models in terms of speed; p-box {\em cdf}-intervals have a closer speed, followed by the {\em fuzzy} models, then by the probabilistic models. In summary, the p-box {\em cdf}-intervals performance is closer to that of the convex models. This means that, the new framework, with minimal overhead, adds up a quantifiable information by imposing tighter bounds on the probability distribution, in a safe and in a tractable manner. Applied computations are tractable because they are exerted on the interval bounds, using interval computations, then results are further projected, linearly, onto the {\em cdf} domain. Empirical evaluations proved that p-box {\em cdf}-intervals have a scalability measure that is close to that of convex models. 
\section{Conclusion and future research direction}
This research proposes a novel constraint domain to reason about data with uncertainty. The key idea is to extend convex models with the notion of p-boxes in order to realize aditional quantifiable information on the data whereabouts. P-Boxes have never been implemented in the CP paradigm, yet they are very good candidates to deal with and reason about uncertainty in the probabilistic paradigm, especially when data is shaping an unknown distribution. 
The case study of the inventory management problem demonstrates that p-box {\em cdf}-intervals can be practically adopted to intuitively envelop the uncertain data found in different modeling aspects with minimum overhead. Evaluation results show that stochastic CPs and probabilistic models have the slowest performance. Fuzzy models proved to have a better performance and their output solutions are characterized to be reliable, i.e. they seek the satisfaction of all possible realizations. Convex models and the p-box {\em cdf}-intervals encapsulate all possible distributions of the solution set in a convex representation. The p-box cdf-intervals framework provides a range of quantiles along with bounds on their data whereabouts.

The introduction of a novel framework to reason about data coupled with uncertainty due to ignorance or based on variability, paves the way to many fruitful research directions. We can list many in: studying models having variables following dependent probability distributions, exploring different search techniques, revisiting the framework within a dynamically changing environment, generalizing the framework to deal with all types of uncertainty by considering together vagueness and dynamicity, and last but not least applying the model to a variety of large scale optimization problems which target real-life engineering and management applications.
\bibliographystyle{acmtrans}
\bibliography{pboxcdfintervals}
\label{lastpage}
\end{document}